# Automatic Discourse Segmentation: Review and Perspectives

*Iria da Cunha*
*Laboratoire Informatique d'Avignon, UAPV, France*
*Institut Universitari de Lingüística Aplicada, UPF, Spain*
iria.dacunha@upf.edu

*Juan-Manuel Torres-Moreno*
*Laboratoire Informatique d'Avignon, UAPV, France*
juan-manuel.torres@univ-avignon.fr

*Abstract*—Multilingual discourse parsing is a very prominent research topic. The first stage for discourse parsing is discourse segmentation. The study reported in this article addresses a review of two on-line available discourse segmenters (for English and Portuguese). We evaluate the possibility of developing similar discourse segmenters for Spanish, French and African languages.

## 1. INTRODUCTION

Discourse analysis has been a widely studied topic since the sixties. There are several relevant discourse theories, most of them language-independent, as for example, the Grice's Maxims [1], the Cohesion Theory [2], the Argumentation Theory [3] or the Theory of Relevance [4], belong others. Nevertheless, the most used discourse theory in computational linguistics is the Rhetorical Structure Theory (RST) [5].

Nowadays automatic discourse parsing is a very prominent research topic, since it is useful to text generation, automatic summarization, automatic translation, textual analysis, information extraction, etc. There are several discourse parsers for English [6, 7], Japanese [8] and Brazilian Portuguese [9, 10]. Most of them use the framework of the RST.

* This work has been financed by a postdoctoral grant of the Spanish Ministry of Science and Innovation (National Program for Mobility of Research Human Resources; National Plan of Scientific Research, Development and Innovation 2008-2011).

The first stage for discourse parsing is discourse segmentation, so a discourse segmenter is necessary to develop a discourse parser. Moreover, a discourse segmenter is useful to tasks involving human discourse annotation, since it allows the annotators to perform their analysis starting form the same automatic segmentation. There are available on-line discourse segmenters for English [11] and Brazilian Portuguese [12]. To our knowledge, there are not discourse parsers and discourse segmenters for Spanish, French or African languages.

In this work, a review of two on-line available discourse segmenters (for English and Portuguese) is presented. Moreover, we evaluate the possibility of developing similar discourse segmenters for Spanish, French and African languages.

In Section 2, both segmenters are showed. In Section 3, the possibilities of developing discourses segmenters in other languages, including African languages, are considered. In Section 4, conclusions are presented.

## 2. Discourse Segmenters

Both discourse segmenters we present here are based on the RST. The RST is a descriptive theory for textual organization that has been proven to be very useful to describe a document by characterizing its structure with relations maintained among its discursive or rhetorical elements (e.g., Circumstance, Elaboration, Motivation, Evidence, Justification, Cause, Purpose, Antithesis, Condition). It is also based on some premises: hierarchy functionality, text structure communicative role and oriented discourse structure predominance. RST determines a set of relations among the discursive units of texts. As a rule, one of the units is the government (*nucleus*), while the other one (*satellite*) provides some rhetorical information about it. This is the more usual structural model between these two units (almost always adjacent units, although there are some exceptions). These relations are named "nuclear" relations. In the case of relations with more than one central unit with regard to author's purposes, the relation is named "multinuclear" and a coordinated relation is established.

As [11] state, "Discourse segmentation is the process of decomposing discourse into elementary discourse units (EDUs), which may be simple sentences or clauses in a complex sentence, and from which discourse trees are constructed". Reference [13] is a manual for EDU segmentation that follows the principles of the RST. These EDUs constitute the satellites or the nuclei of the RST. Fig. 1 shows an example of a discourse tree.

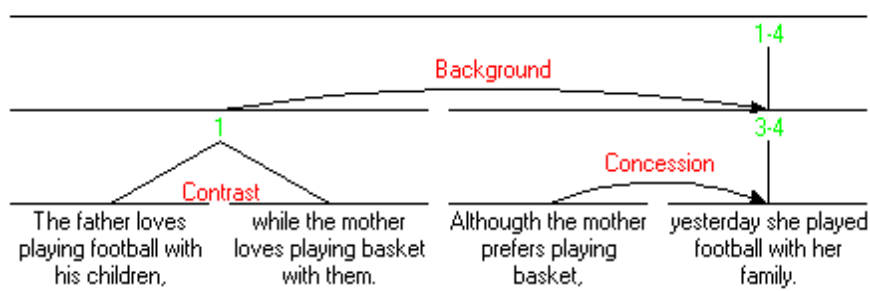

Figure 1. Example of a RST discourse tree

In [11], the authors have developed SLSeg, a Syntactic and Lexical-Based Discourse Segmenter for English, based on the RST. Users can download the system in the following URL: http://www.sfu.ca/~mtaboada/research/SLSeg.html. With regard to EDU segmentation, they revised some questions from [13]. As they indicate: "Some specifications were made so that we would be able to clearly differentiate syntactic and discursive levels. In this work, we consider that EDUs must include a finite verb (that is, they have to constitute a sentence or a clause) and must show, strictly speaking, a rhetorical relation." Most of these specifications are very similar of those included in [14], where the discourse structure of Spanish and Basque texts is studied. The approach of [11] uses a set of 12 lexical and syntactic rules for insert segment discourse boundaries in English texts. Two morphosyntactic parsers are compared in order to evaluate their effect on segmentation quality: Charniak [15] and Stanford [16]. Finally, they evaluate SLSeg comparing it with another discourse segmenter [17] and with some baselines. They obtain very good results.

In [12], the authors have developed a discourse segmenter for Brazilian Portuguese. The system can be used on-line in the following URL: http://www.nilc.icmc.usp.br/~erick/segmenter/. With regard to EDU segmentation, they follow the guidelines included in [13]. The approach of [12] use a set of 12 lexical and syntactic rules for insert segment discourse boundaries in Brazilian Portuguese texts. They use the output of the morphosyntactic parser PALAVRAS [18]. The authors of [12] explain the discourse segmentation process developed (that will be included into DiZer, the discourse parser of [9, 10]). Nevertheless, they do not carry out an explicit evaluation of the tool, as [11] do.

### 3. PERSPECTIVES OF DEVELOPING DISCOURSE SEGMENTERS IN OTHER LANGUAGES

English is a Germanic language and Portuguese is a Romance language. Nevertheless, both languages have common characteristics, as they share a similar word conception, they use the Latin alphabet, they employ sentences and clauses, etc. In order to develop discourse segmenters for other languages, the first stage is to evaluate if it is possible to employ similar strategies to the ones used in [11] and [12]. In this section, we assess, *grosso modo*, this possibility, both for Romance languages (Spanish and French) and African languages.

### *3.1 Perspectives for Spanish and French*

As explained it Section 2, the main strategy to develop a discourse segmenter includes some phases:
1. Constitution of a textual corpus.
2. RST discourse annotation of the corpus.
3. Manual analysis of the lexical and syntactic elements of the corpus that could be considered as segment boundaries.
4. Implementation of lexical and syntactic discourse segmentation rules.
5. Use of the output of a morphosyntactic parser as input for the segmentation rules.
6. Evaluation of the developed discourse segmenter.

We are conscious of the limitations of the rule-based computational resources (language-dependent, need to collaborate with linguists, etc.), but there are not statistical studies on this topic, so our strategy is to adapt to other languages the methodology that [11] and [12] use.

At the Laboratoire Informatique d'Avignon, we work on a project that aims to develop a Spanish discourse parser. The first stage of this project is to implement a discourse segementer for this language. For the phases 1-5 we do not have any problem: 1) we have a medical corpus, 2) we have the RSTTool [19] for the RST discourse annotation, 3) we have performed the manual analysis, 4) we have implemented the lexical and syntactic discourse segmentation rules and 5) we have used the Spanish morphosyntactic parser Freeling [20]. Nevertheless, in the phase 6, we have found a limitation: at present there is not another Spanish discourse segmenter, so we can not evaluate our segmenter comparing it with another one, as [11] do. The provisional solution is to evaluate the segmenter comparing it with some baselines.

The lexical and syntactic discourse segmentation rules are mainly based on discourse markers, conjunctions, verbal forms and punctuation marks. For example, one of the rules indicates that an EDU always has to include a finite verb. Thus, the sentence 1a would be separated into two EDUS, while the sentence 1b would constitute a single EDU.

     1a. [The hospital is adequate to adults,]$_{EDU1}$ [but children can use it as well.]$_{EDU2}$

     1b. [The hospital is adequate to adults, as well to children.]$_{EDU1}$

Another rule indicates that if a sentence includes a discourse marker (as, for example, "because", "nevertheless", "if", "but", etc.) it has to be separated into two EDUs (if both include a finite verb), as sentence 2 shows.

     2. [The emergency services of this hospital are very efficient,]$_{EDU1}$ [because patients are always seen immediately.]$_{EDU2}$

If the Spanish segmenter obtains good results, we would try to use the same methodology in order to develop the French discourse segmenter, as both are Romance languages and they have a similar structure. There are several available French textual corpora and some morphosyntactic parsers for this language (see, for example, [21, 22]). Once again, there is not another French discourse segmenter for the evaluation.

## *3.2 Perspectives for African languages*

In Africa there are approximately 2000 spoken languages. They are usually divided into six major linguistic families: Afroasiatic, Nilo-Saharan, Niger-Congo, Khoe, Austronesian and Indo-European. Although there are so many African languages, most of them are basically spoken and they do not have many textual resources. As [23] assess: "Ces corpus textuels sont inexistants dans la plupart de ces pays du fait, précisement, de leur tradition orale". Moreover, these languages have very few electronic resources, that is, they are not very computerized languages. Some researchers are conscious of this situation, and they are working to solve this problem [23, 24, 25]. An example is the work of [23], who have constituted automatically a corpora for the Somali language with 3 millions of words using Internet sites (as newspapers sites, cultural associations sites, etc.). Moreover, they present a system of automatic text generation and an automatic translator of Somali.

With regard to discourse segmentation, there are not studies for African languages. Maybe, the proposed methodology could be useful to develop this segmenter:

1. As we have mention before, there are some African groups researching on textual corpora, although all the African languages are not covered by these studies. The available textual corpora could be used for this task.

2. To carry out the RST discourse annotation, the RSTTool can be used for the African languages with Latin alphabet.

3. For the manual analysis, the principal needed resource is a research team knowing the RST principles. Nevertheless, it is also necessary some background regarding the topic of discourse markers in African languages, as they are necessary in order to develop the discourse segmenter rules. As [26] states:

   "Seminal works […] have led to a growing body of literature on discourse markers in recent decades, with an increasingly diverse typological range of languages from different areal and genetic classifications [...]. African languages, however, remain largely unrepresented in the study of discourse markers in particular as well as sociolinguistic variation in general, as the bulk of sociolinguistic studies of African languages focus on topics such as ethnopragmatics, language contact and language death, language policy and other issues of multilingualism, and code-switching."

4. For the implementation of the segmentation rules, a programming language is needed (as for example, Perl).

5. A morphosyntactic parser is necessary to use as input for the rules. To our knowledge, there is no suitable morphosyntactic parsers for African languages. This fact constitutes one of the main problems of the methodology.

6. As in the case of Spanish and French, it would not be possible to use another discourse segmenter for the evaluation.

## 4. Conclusions

In this work, we have presented two relevant available discourse segmenters, for English and for Brazilian Portuguese. Moreover, we have shown that both use a very similar methodology, although English is a Germanic language and Brazilian Portuguese is a Romance language. This methodology could be used in order to develop discourse segmenters for other languages. In the case of languages with many linguistic and computational resources, as for example Spanish and French, the task would be easier than in the case of languages with very few resources, as African languages. The common problem is the lack of other discourse segmenters to compare their results and to carry out the evaluation. The main problems that we have found to develop this tool for the African languages are the lack of morphosyntactic parsers and the lack of background with regard to the topic of discourse markers. As [27] wonders: "Mais que se passe-t-il dans les langues des pays en développement lorsque font irruption ces nouvelles sciences et techiques ? Sont-elles capables de faire face ? Ont-elles la capacité d'exprimer cette modernité exogène ?". We think the answer to these questions is yes. If the African research community continues with the investigation on Natural Language Processing (NLP), we are sure that African languages will have linguistic and computational resources in the future, and these will be very beneficial for many cultural, educational and politic domains, among others.